# A New Paradigm for Fault-Tolerant Computing with Interconnect Crosstalks

Naveen Kumar Macha[1*], Bhavana Tejaswini Repalle[1], Sandeep Geedipally[1], Rafael Rios[2], Mostafizur Rahman[1#]
E-mail: nmhw9@mail.umkc.edu[1*], rahmanmo@umkc.edu[1#]

*Abstract*— The CMOS integrated chips at advanced technology nodes are becoming more vulnerable to various sources of faults like manufacturing imprecisions, variations, aging, etc. Additionally, the intentional fault attacks (e.g., high power microwave, cybersecurity threats, etc.) and environmental effects (i.e., radiation) also pose reliability threats to integrated circuits. Though the traditional hardware redundancy-based techniques like Triple Modular Redundancy (TMR), Quadded (QL) Logic etc. mitigate the risk to some extent, they add huge hardware overhead and are not very effective. Truly polymorphic circuits that are inherently capable of achieving multiple functionalities in a limited footprint could enhance the fault-resilience/recovery of the circuits with limited overhead. We demonstrate a novel crosstalk logic based polymorphic circuit approach to achieve compact and efficient fault resilient circuits. We show a range of polymorphic primitive gates and their usage in a functional unit. The functional unit is a single arithmetic circuit that is capable of delivering Multiplication/Sorting/Addition output depending on the control inputs. Using such polymorphic computing units in an ALU would imply that a correct path for functional output is possible even when 2/3rd of the ALU is damaged. Our comparison results with respect to existing polymorphic techniques and CMOS reveal 28% and 62% reduction in transistor count respectively for the same functionalities. In conjunction with fault detection algorithms, the proposed polymorphic circuit concept can be transformative for fault tolerant circuit design directions with minimum overhead.

*Keywords—Fault tolerance, nano-computing, crosstalk, interconnect interference, polymorphic, reconfigurable computing*

## I. INTRODUCTION

As scaling of technology nodes go below 10nm scale, hard and soft errors due to process imprecision, variation, and aging are adversely affecting the yield and reliability of ICs. Fault tolerant circuits can help in mitigating the concerns and increase reliability. A truly fault resilient circuit scheme can also gracefully recover from run-time faults such as those that incur due to radiation, high-power microwave, and cyber threats. Traditional approach for fault tolerance has been concentrated on redundancy based circuits such as CMOS circuit Multiplexing [1], Triple Modular Redundancy (TMR) and its generalized extension N-tuple Modular Redundancy (NMR) [2], Triplicated Interwoven Redundancy and its generalized extension N-tuple Interwoven redundancy (NIR)[3], and Quadded Logic [4] etc. The need for duplication of logic in all the above approaches/schemes results in large overhead. A more recent approach for fault tolerance looks at circuit level reconfigurability/polymorphism to achieve multiple functionalities with a single logic block. The motivation for such scheme is illustrated in Fig. 1. When a single gate (Fig. 1(i)) is affected by a fault and malfunction, another working gate (the NAND in this example) can be used to perform both functionalities. The gate level reconfigurability concept can be extended to module and system level also as depicted in Fig. 1(ii). Although such polymorphic concepts are enabling, a scalable CMOS alternative paradigm to achieve this is lacking. Existing approaches either rely on environmental control variables such as light, temperature etc. [5] or require new exotic switches [6][7] that are yet to mature.

In this paper, we show a novel solution for polymorphic circuits using interconnect crosstalks. Previously, we demonstrated how interference between two signal carrying metal nano-lines can be engineered for logic operation [8]. Here, we show how the same principle can be extended for reconfiguration. For operation, the transition of signals on input metal lines (including polymorphic control signal) called as aggressor nets induce a resultant summation charge on output metal line called as victim net through capacitive couplings. This induced signal serves as an intermediate signal to control thresholding devices like pass-transistor or an inverter to get the desired logic output. To achieve polymorphic behavior, the victim net is influenced/biased by a control aggressor, which switches the circuit behavior to a different logic type. In this paper, we show how polymorphism allows reconfiguration of basic gates such as NAND-NOR, AND-OR, AOI-OAI and functional units such as Mutlipler-Sorter-Adder. We also present a comparison with CMOS and other available technologies. Our results indicate at-least 62% reduction in transistor count compared to CMOS and 28% reduction compared to other polymorphic approaches for the same functionality. Also, we introduce the new polymorphic circuits based fault tolerance concept applicable from gate-level to module and system level. Additionally, we present high-level fault discovery and fault-recovery routines for system level utilization of polymorphic-crosstalk circuits.

The rest of the paper is organized as follows: Section.II describes the crosstalk computing fundamentals, Section.III presents polymorphic gates implementation in Crosstalk fabric for fault resilience and a large functional unit example. Section IV compares with other polymorphic circuits in literature. Finally, Section V presents the conclusion.

[1]Authors are with Computer Science and Electrical Engineering Department at University of Missouri Kansas City
[2]R. Rios is an independent consultant, formerly with Intel and AMD

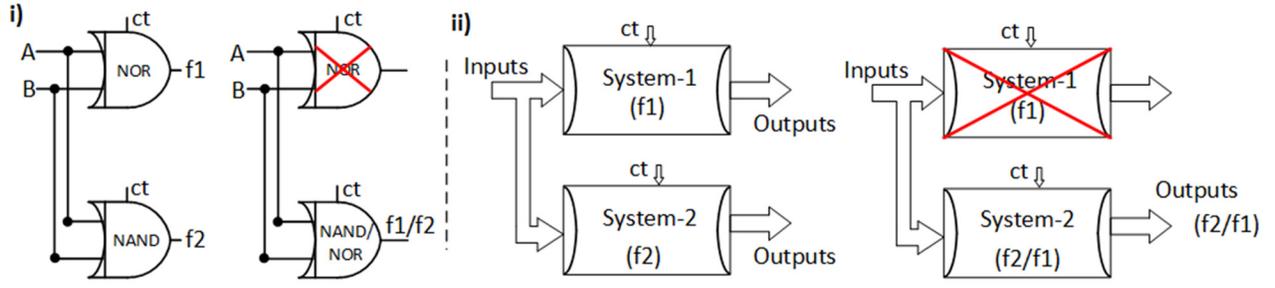

Fig.1. Polymorphic/Re-configurable circuit based Fault Tolerance concept, i) Gate-level, ii) System-level

## II. CROSSTALK COMPUTING FUNDAMENTALS

The logic computation in crosstalk computing fabric happens in metals lines, coupled with accurate control and reconstruction of signals in transistors. We have introduced the crosstalk computing concept in [8]. The crosstalk-logic can implement efficiently both linear logic functions (e.g., AND, OR etc.) and non-linear logic functions (e.g. XOR). The primary principle for logic computation is through deterministic charge induction in the output node. Fig. 2(i.a) shows a NAND gate. Here, the transition of the signals on two adjacent aggressor metal lines ($Ag1$ and $Ag2$) induces a resultant summation charge/voltage on victim metal line ($Vi$) through capacitive coupling. Since this phenomenon follows the charge conservation principle the victim node voltage is deterministic in nature, therefore it can be stated that the signal induced on victim net posses the information about signals on two aggressor nets, and its magnitude depends upon the coupling strength between the aggressors and victim net. This coupling capacitance is inversely proportional to the distance of separation of metal lines and directly proportional to the relative permittivity of the dielectric and lateral area of metal lines (which is length x vertical thickness of metal lines). Tuning the coupling capacitance values using the variables mentioned above provides the engineering freedom to tailor the induced summation signal to the specific logic implementation or as an intermediate control signal for further logic. For example, OR gate requires strong coupling than AND gate, which can be achieved by tuning the dimensions and high-k dielectric material choices. In Fig. 2(i.a), a discharge transistor driven by $Dis$ signal and an inverter are connected to $Vi$ net as shown in the figure. The Crosstalk logic operates in two states, logic evaluation state (ES) and discharge state (DS). During ES, the rise transitions on aggressor nets induce proportionate linear summation voltage on Vi (through couplings) which is connected to a CMOS inverter acting as a threshold function. During discharge state (enabled by $Dis$ signal) floating victim node is shorted to ground through discharge transistor, this ensures correct logic operation during next logic evaluation state (ES) by clearing off the value from the previous logic operation. The simulation response of the designed NAND gate is shown in Fig.2.(i.b). For required capacitance representation, we will use crosstalk-margin function $CT_M(C_{ND})$, which specifies that the inverter of the crosstalk polymorphic logic gate flips its state only when victim node sees the input transitions through the total coupling greater than or equal to $C$. For example, NAND CT-margin function is $CT_M(2C_{ND})$, which states that inverter flips the state only when victim node sees the input transitions through total coupling greater than or equal to $2C_{ND}$, i.e. when both inputs are high.

A more complex logic implementation is shown in Fig. 2(ii.a) through an AOI21 circuit. Logic expression of AOI21, $(AB+C)'$, evaluates to 0 when either $AB$ or $C$, or both are 1. That means the output is biased towards the input $C$ i.e., irrespective of $A$ and $B$ values, the output is 1 when $C$ is 1. Therefore, in Fig.2(ii.a), input $C$ has the coupling $2C_{AO}$, whereas, $A$ and $B$ have $C_{AO}$ capacitance. The margin function for this gate is $CT_M(2C_{AO})$. The response of the circuit is shown in Fig.3(ii.b). A complete list of fundamental and complex polymorphic gates with Crosstalk computing can be found in [9].

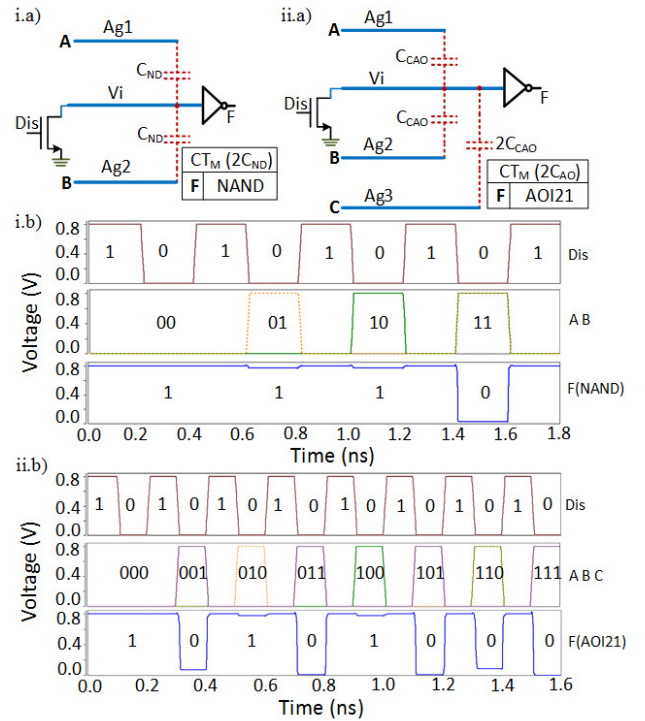

Fig.2. Crosstalk Logic Gates: i(a) NAND circuit, i(b) NAND simulation response, ii(a) AOI21 circuit, ii(b) AOI21 simulation response

## III. POLYMORPHISM FOR FAULT TOLERANCE

### A. Basic Gate Level Polymorphism

The polymorphic logic gates exhibit multiple logic behaviors by altering a control variable, as a result, increases the logic expressibility of a circuit. A wide range of polymorphic gates can be implemented using crosstalk circuit techniques, out of which, we show here the circuit reconfigurability between AND/OR, OA21/AO21, AND3/AO21 and AO21/OR3. These circuits switch the logic behavior by using an additional control aggressor. Fig.3(i) shows the crosstalk-polymorphic AND/OR circuit and its response graph. As shown in the circuit diagram, inputs ($A$ and $B$) and control aggressor ($Ct$) has the coupling $C_{PA}$ (the coupling capacitance values are detailed in Table.1). $F_I$ stage gives the inverting function (NAND/NOR) response and $F$ stage gives non-inverting function (AND/OR). A table adjacent to circuit diagram lists the margin function and the circuit operating modes. The margin function for AND/OR cell is $CT_M (2C_{PA})$. When control $Ct=0$ it operates as AND, whereas, when $Ct=1$ the $Ct$ aggressor (Ag3) augments charge on to $Vi$ net through the coupling capacitance $C_{PA}$, hence, following the function $CT_M(2C_{PA})$ the cell is now biased to operate as an OR gate, therefore, the transition of either $A$ or $B$ is now sufficient to flip the inverter. The same response can be observed in the simulation plots shown in the Fig.3(i.b), the first panel shows the discharge ($Dis$) and control ($Ct$) signals, 2nd panel shows the input combinations fed through $A$ and $B$, and 3rd panel shows the response at the stage $F$. It can be observed that the circuit responds as AND when Ct=0 for first four input combinations (00 to 11), whereas, it responds as OR when Ct=1 during next four input combinations (00 to 11).

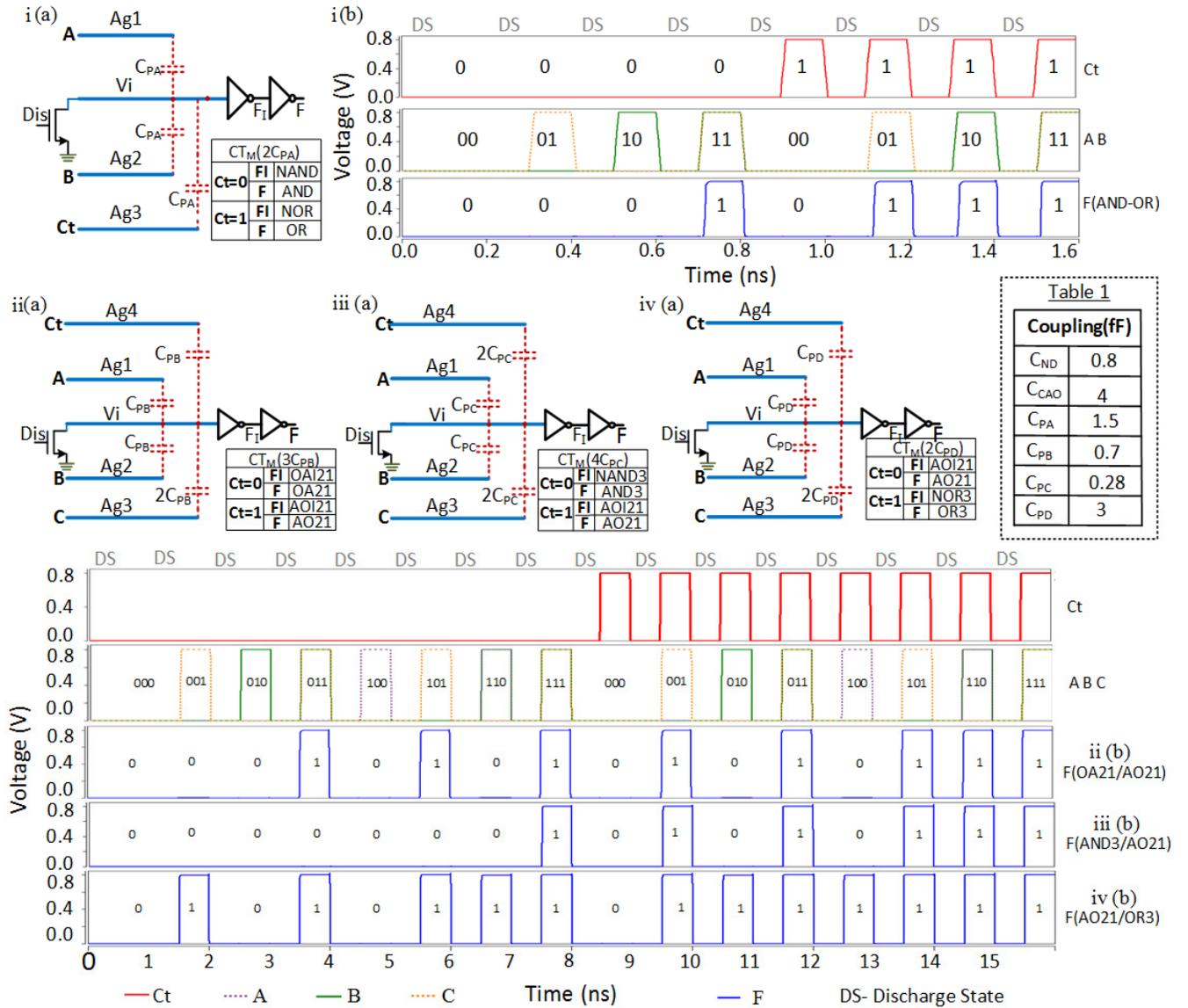

Fig.3. Crosstalk Polymorphic logic gates and their simulation responses: i(a) AND/OR circuit, i(b) AND/OR simulation; ii(a) OA21/AO21 circuit, ii(b) OA21/AO21 simulation; iii(a) AND3/AO21 circuit, iii(b) AND3/AO21 simulation; iv(a) AO21/OR3 circuit, iv(b) AO21/OR3 simulation

The next three circuits depicted in Fig.3(ii.a)-3(iv.a) implement 3 variable polymorphic functions OA21/AO21, AND3/AO21 and AO21/OR3. The simulation responses of these circuits are presented in the waveforms below (Fig.3(ii.b)-3(iv.b)). The first and second panels show the inputs $Ct$, and $A$, $B$ and $C$ respectively, while the panels below are the responses of the circuits in Fig.3(ii.a) to Fig.3(iv.a) respectively. For OA21/AO21 circuit (Fig.3(ii.a)), aggressors $A$, $B$, and $Ct$ are given $C_{PB}$ coupling, whereas input $C$ is given $2C_{PB}$, the margin function is $CT_M$ ($3C_{PB}$). When control $Ct=0$ it operates as OA21, whereas, when $Ct=1$ the $Ct$ aggressor ($Ag4$) augments charge through the coupling capacitance $C_{PB}$, hence, following the function $CT_M(3C_{PB})$ the cell is now biased to operate as AO21. The same response can be observed in the simulation graph (4th panel), the circuit responds as OA21 when $Ct=0$ for first eight input combinations (000 to 111), whereas, it responds as AO21 when $Ct=1$ for next eight combinations (000 to 111). Similarly, Fig.3(iii.a) depicts AND3/AO21 circuit, where, $A$ and $B$ are given $C_{PC}$ coupling, while $Ct$ and $C$ are given $2C_{PC}$ coupling, and the margin function here is $CT_M$ ($4C_{PC}$), therefore, the circuit responds as AND3 (4th panel) for all input combinations when $Ct=0$, whereas, it responds as AO21 when $Ct=1$. Similarly, for the AO21/OR3 circuit in Fig.3(iv.a) the coupling choices for $A$, $B$ and $Ct$ are $C_{PD}$, and for $C$ it is $2C_{PD}$. Following the margin function $CT_M$ ($2C_{PD}$), the circuit behaves as AO21 when $Ct=0$ for first 8 input combinations (000 to 111), while it behaves as OR3 when $Ct=1$ for next 8 input combinations (000 to 111). By employing such compact and efficient polymorphic logic gates in the circuits, in the event of a fault occurrence in some portions of the circuit, the unaffected logic gates can be morphed to implement the damaged functionality, thus, it could pave ways to a new paradigm of fault tolerance which is based on polymorphism at gate-level.

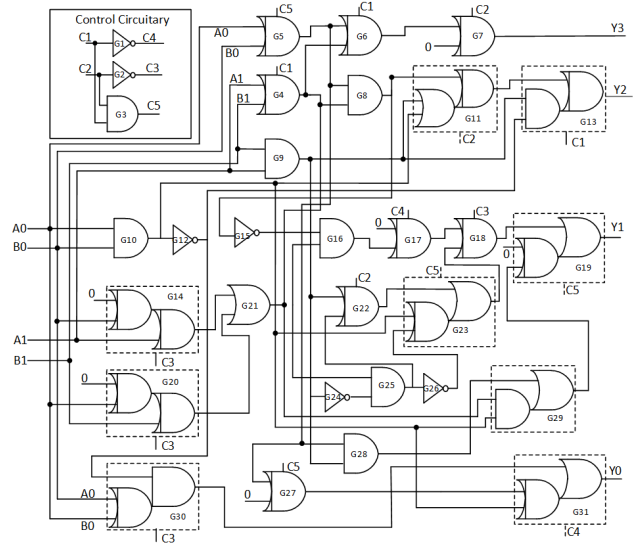

Fig.4. Crosstalk Polymorphic Multiplier/Adder/Sorter circuit

### B. Block Level Polymorphism

This section demonstrates the block level polymorphism using a circuit example of 2-bit multiplier-sorter-adder (Fig.4) which is implemented using the polymorphic gates discussed above. The circuit uses 31 gates in total, out of which 25 are crosstalk gates, and 6 are inverters. 16 out of 25 crosstalk gates are polymorphic gates which are efficiently employed to switch the circuit between the multiplier, sorter and adder operations using two control signals (*C1* and *C2*). The inset figure shows the control circuitry (*C1-C5*). Fig.5 shows the simulation response of the circuit, different operation modes of the circuit are annotated on top, which are, Multiplier (M), Sorter (S), and Adder (A). The first panel shows *Dis* signal, *Dis=1* is the discharge state (DS) and *Dis=0* is the logic evaluation state. The second panel shows the control signals *C1* and *C2* whose values as 01, 11 and 10 corresponds to multiplier, sorter and adder operations. Third and fourth panels show the 2-bit inputs *A[1:0]* and *B[1:0]* respectively, the following panels show the 4-bit response of the circuit *Y[3:0]*. The circuit is operated alternately in the multiplier, sorter, and adder modes, and in each set of this modes, common input values are fed through *A1A0* and *B1B0* which effectively demonstrates the transformation of the circuit in accordance with the control signals. For example, for the first input combinations, 11 and 10, the multiplier operation gives 0110 as output while the succeeding sorter and adder operations give 1110 and 0101 outputs respectively. Similarly, for the second inputs 10 and 01, M, S, and A operations give 0010, 1100 and 0011 outputs respectively. In similar fashion, few other combinations are shown in the next stages. The circuit consumes only 155 transistors in total. Such polymorphic circuits can be employed for the fault tolerance at the block level. For example, as shown in the Fig.6, Multiplier, Sorter and Adder operations can be implemented in independent blocks which also possess the dormant other two operations. During the event of fault detection in one of the blocks, the other blocks can be reconfigured and multiplexed to achieve the correct output. The polymorphic blocks can be also used with traditional voter based [2] fault resiliency techniques.

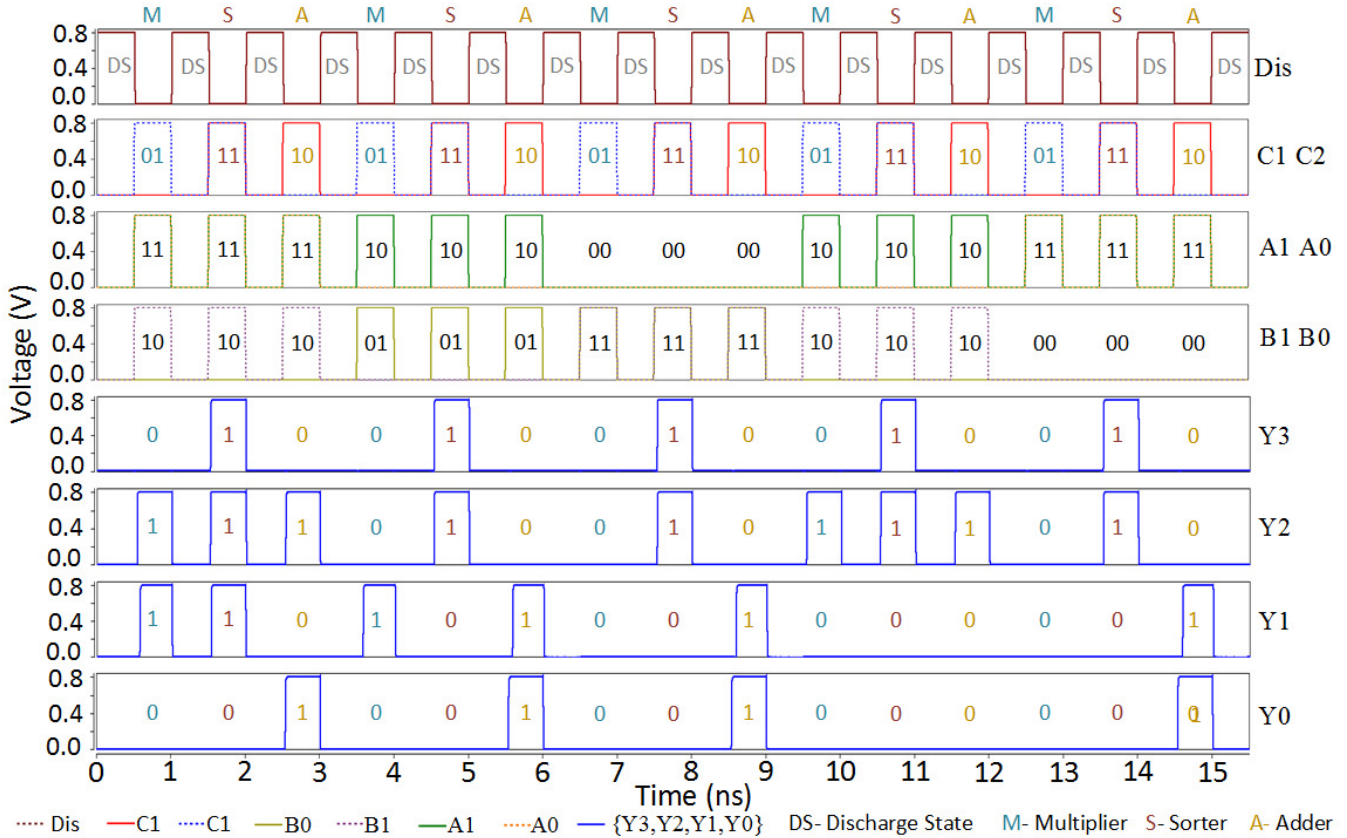

Fig.5. Crosstalk Polymorphic Multiplier/Adder/Sorter circuit simulation response

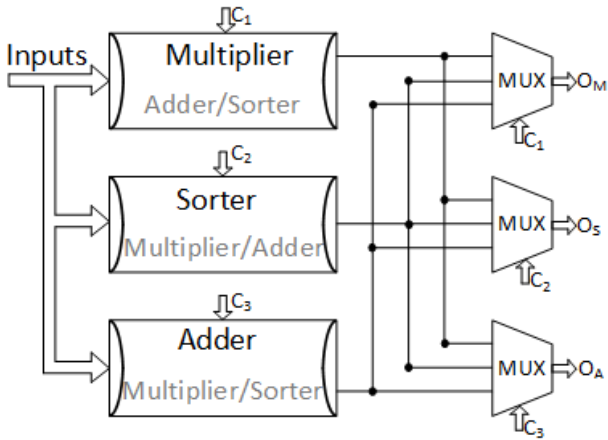

Fig.6. Block-Level Polymorphic Fault Tolerant scheme

## C. System Level Polymorphism

In Fig.7, we introduce the concept of hardware-software based fault detection and recovery scheme that can fully utilize the polymorphic circuits to recover from faults at run-time. Here, polymorphic circuit blocks are deployed first and periodically monitored during operation for correctness and recovery. First, a block is configured for one operation and known set of inputs are driven to check the functional correctness. If the correct operation is registered, the block and operation is registered in a lookup table. Similarly, all blocks and relevant functionalities are checked and their information are stored in the lookup table. Upon fault detection in one of the blocks, the Software/Assembler will look for alternative blocks in the lookup table, and re-route and reconfigure blocks accordingly to achieve correct results.

**Fault Discovery**

*//Assuming n different functional blocks are available and each block can be configured to achieve m different functionalities*

*Step 1: Configure block1 for functionality 1 by asserting configuration bits*

*Step 2: Drive known set of inputs for functional verification*

*Step 3: Check outputs for correctness*

*Step 4: If outputs are correct/incorrect, mark block1 functionality 1 as correct/incorrect, configure block1 for functionality 2 … functionality m and repeat from Step 2.*

*Step 5: Repeat Steps 1 to 4 for all computing blocks to discover working functions*

**Fault Recovery**

*Step 1: Run Fault Discovery algorithm to discover all correct computing blocks and their respective functions.*

*Step 2: Operating system stores information about correct blocks and functions in lookup table and generates instructions accordingly*

*Step 3: From incoming instructions, configure bits are generated during instruction decode phase and all blocks are configured*

*Step 4: All output multiplexers get proper selection inputs*

*Step 5: Inputs are driven to computing blocks and outputs observed*

Fig.7. Fault Discovery and Recovery Steps

## IV. COMPARISON & DISCUSSION OF MERITS

The proposed fault tolerant scheme can be implemented using any polymorphic circuit approaches [5-7][10-11]. In order to quantify the efficacy, in this section, we compare the crosstalk polymorphic logic technology with respect to CMOS multiplexer based polymorphic implementation and to a recent approach of ambipolar Si-Nanowire circuits [10][11]. The traditional approach ('CMOS' column in the table) is multiplexer based, where independent stand-alone circuits are designed and selected through a multiplexer, the hardware redundancy in this method is huge. Whereas, in the second approach circuits are constructed using nanowire transistors which are configurable to either n-type or p-type using a control voltage. Limitations of this approach are, the density benefit is limited, additional circuitry is required to swap power rails for pull-up and pull-down networks, device response is not robust, and it imposes the complex manufacturing steps. The other alternate approach for polymorphic circuits is using emerging spintronic devices [6], but they rely on complex information encoding scheme through spin-polarized currents, and bipolar voltages etc., also, they are a significant departure from existing computational, device and circuit paradigm.

Table 2. Comparison with relevant technologies

| | CMOS | Ambipolar NWFET[10] | Crosstalk-Polymorphic |
|---|---|---|---|
| **Computing Principle** | Complementary logic with FETs | Band Structure modification | Control of signal interference |
| **Application for fault tolerant computing** | Circuit duplication (redundancy) and use of multiplexers to select redundant blocks | Polymorphic logic implementation with configurable transistors | Logic reconfiguration by leveraging Crosstalk |
| **Trade-off** | Density, power and performance penalties for redundant blocks | Scalability challenges and limited density benefits | Control overhead vs. Density, Power & Performance benefits |
| Transistor Count Comparison | | | |
| **AND2-OR2** | 18 | 6 | 5 |
| **AO21-OA21** | 22 | 8 | 5 |
| **AND3-AO21** | 22 | 12 | 5 |
| **AO21-OR3** | 22 | 12 | 5 |
| **Multiplier-Sorter-Adder** | 408 | 216 | 155 |

The crosstalk-polymorphic approach compared to other approaches is very compact in implementation, friendly to advanced technology nodes and scalable to the larger polymorphic systems. Also, the working mechanism is simple and reliable. The benefits in performance metrics like, area, power and performance are also best compared to any other approaches. Deliberate and very fast reconfigurability is achievable by using a control signal. The benchmarking of transistors count requirement for basic, complex and cascaded logic cases are given in the table.2 The complex gates listed for the Si-NWFET approach are constructed by cascading polymorphic NAND-NOR, AND-OR gates presented in [10]. The crosstalk-circuit based polymorphic approach consumes fewer transistors than any other approaches. The transistor count comparisons show that crosstalk polymorphic gates show the reduction ranging from 25% to 83% at the cell level. For the multiplier/adder/sorter circuit, our approach show 28% and 62% reduction in transistor count compared to ambipolar SiNWFET and CMOS approaches respectively. Moreover, unlike any other approaches, crosstalk-polymorphic circuits could implement a wide range of complex logic functions in a compact manner [9].

## V. CONCLUSION

We introduced a novel polymorphic circuit approach for fault tolerant computing leveraging interconnect crosstalks. Various polymorphic logic gates including AND-OR, AO21-OA21, AND-AO21, and OR-AO21 and a complex circuit that performs the functionalities of Multiplication/Sorting/Addition with polymorphic gates were shown. Transistor count comparison revealed potential benefits of crosstalk-polymorphic logic; for the same complex circuit implementation, the transistor count was found to be 155 vs. 408 of CMOS. We also presented an approach for run-time system level fault detection and recovery. The proposed work sets new pathways for fault-tolerant computing and can be transformative for reliable integrated circuits in future.

REFERENCES

[1] 1. J. von Neumann, "Probabilistic Logics and the Synthesis of Reliable Organisms from Unreliable Components," AutomataStudies, C.E. Shannon and J. McCarthy, eds., Princeton Univ. Press, 1956, pp.

[2] Dubrova E, Fault-tolerant design, Springer, 2013.

[3] Jie Han, J. Gao, P. Jonker, Yan Qi and J. A. B. Fortes, "Toward hardware-redundant, fault-tolerant logic for nanoelectronics," in IEEE Design & Test of Computers, vol. 22, no. 4, pp. 328-339, July-Aug. 2005.

[4] J. Han, E. Leung, L. Liu, F. Lombardi, "A fault-tolerant technique using quadded logic and quadded transistors," IEEE Transactions on VLSI Systems, vol. 23, no. 8, pp. 1562-1566, August 2015.

[5] A. Stoica, R. Zebulum, and D. Keymeulen, "Polymorphic Electronics," Evolvable Syst. From Biol. to Hardw., vol. 2210, pp. 291–302, 2001.

[6] S. Rakheja and N. Kani, "Polymorphic spintronic logic gates for hardware security primitives — Device design and performance benchmarking," 2017 IEEE/ACM International Symposium on Nanoscale Architectures (NANOARCH), Newport, RI, 2017, pp. 131-132.

[7] Yu, W. J., Kim, U. J., Kang, B. R., Lee, I. H., Lee, E. H., Lee, Y. H, "Multifunctional logic circuit using ambipolar carbon nanotube transistor," Proc. SPIE 7399, 739906 (2009).

[8] Naveen kumar Macha, et al., "A New Concept for Computing Using Interconnect Crosstalks," Rebooting Computing (ICRC), 2017 IEEE International Conference, Washington, DC, USA, December 2017.

[9] Naveen kumar Macha, Sandeep Geedipally, Bhavana Tejaswee Repalle, Md Arif Iqbal, Wafi Danesh, Mostafizur Rahman "Crosstalk based Fine-Grained Reconfiguration Techniques for Polymorphic Circuits," IEEE/ACM NANOARCH 2018 (submitted)

[10] M. De Marchi et al., "Configurable logic gates using polarity controlled silicon nanowire gate-all-around FETs," IEEE Electron Device Lett., vol.35, no. 8, pp. 880–882, 2014.

[11] J. Zhang, P. E. Gaillardon, and G. De Micheli, "Dual-threshold-voltage configurable circuits with three-independent-gate silicon nanowire FETs," Proc. - IEEE Int. Symp. Circuits Syst., pp. 2111–2114, 2013.